\documentclass[11pt]{article}
\usepackage{hyperref}
\usepackage{amsmath,amsthm,amssymb}

\usepackage[small,bf]{caption}

\usepackage{graphicx}

\hypersetup{colorlinks=true, linkcolor=black, citecolor=black,%
pdftitle={Tight Bounds on Information Dissemination in Sparse Mobile Networks},%
pdfauthor={Alberto Pettarin, Andrea Pietracaprina, Geppino Pucci, and Eli Upfal}}

\newcommand{\newterm}[1]{\emph{#1}}

\newcommand{\bigO}[1]{O\left(#1\right)}
\newcommand{\bigOt}[1]{\tilde{O}\left(#1\right)}
\newcommand{\bigTh}[1]{\Theta\left(#1\right)}
\newcommand{\bigTht}[1]{\tilde{\Theta}\left(#1\right)}
\newcommand{\bigOm}[1]{\Omega\left(#1\right)}

\newcommand{\card}[1]{\left|#1\right|}

\newcommand{\dist}[1]{||#1||}
\newcommand{\prob}[1]{\mathrm{Pr}\left({#1}\right)}
\newcommand{\expe}[1]{\mathrm{E}\left[{#1}\right]}
\newcommand{\vari}[1]{\mathrm{Var}\left({#1}\right)}

\newcommand{\lis}[1]{\left\langle #1 \right\rangle}

\newcommand{\bt}{T_{\mathrm{B}}}
\newcommand{\gt}{T_{\mathrm{G}}}
\newcommand{\ct}{T_{\mathrm{C}}}

\newcommand{\B}{\mathcal{B}}
\newcommand{\I}{\mathcal{I}}

\newcommand{\R}{\mathbb{R}}
\newcommand{\Z}{\mathbb{Z}}
\newcommand{\N}{\mathbb{N}}
\newcommand{\grid}{\mathcal{G}_n}
\newcommand{\point}[1]{#1} %
\newcommand{\mes}[2]{M_{#1}(#2)} %

\newcommand{\tf}[1]{\small{#1}}

\newtheorem{defi}{Definition}
\newtheorem{theo}{Theorem}
\newtheorem{lemm}{Lemma}
\newtheorem{coro}{Corollary}

\newenvironment{proofof}[1]{\begin{trivlist} 
                         \item[] {\it Proof of #1:}}{\hfill $\Box$
                       \end{trivlist}}

\begin{document}
\title{Tight Bounds on Information Dissemination\\ in Sparse Mobile Networks%
	\thanks{Support for the first three authors was provided, in part, by MIUR of Italy
	under project AlgoDEEP, and by the University of
	Padova under the Strategic Project STPD08JA32 and Project
	CPDA099949/09.
	This work was done while the first author was visiting the Department of Computer Science of Brown University,
	partially supported by ``Fondazione Ing.~Aldo Gini'', Padova, Italy.}%
}

\author{
Alberto Pettarin\and Andrea Pietracaprina\and Geppino Pucci\\
	\tf{Department of Information Engineering,} %
	\tf{University of Padova}\\
	\tf{\texttt{\{pettarin,capri,geppo\}$@$dei.unipd.it}}
\and%
Eli Upfal\\
	\tf{Department of Computer Science,} %
	\tf{Brown University}\\
	\tf{\texttt{eli$@$cs.brown.edu}}
}

\date{}
\maketitle{}

\begin{abstract}
Motivated by the growing interest in mobile systems,
we study the dynamics of information dissemination
between agents moving independently on a plane.
Formally, we consider $k$ mobile agents
performing independent random walks on an $n$-node grid.
At time $0$, each agent is located
at a random node of the grid and one agent has a rumor.
The spread of the rumor is governed by
a dynamic communication graph process $\{G_t(r)~|~t\geq 0\}$,
where two agents are connected by an edge in $G_t(r)$ iff
their distance at time $t$ is within their transmission radius $r$.
Modeling the physical reality
that the speed of radio transmission
is much faster than the motion of the agents,
we assume that the rumor can travel
throughout a connected component of $G_t$
before the graph is altered by the motion.
We study the \emph{broadcast time} $\bt$ of the system,
which is the time it takes for all agents to know the rumor.
We focus on the sparse case
(below the percolation point $r_c \approx \sqrt{n/k}$)
where, with high probability,
no connected component in $G_t$ has more than
a logarithmic number of agents
and the broadcast time is dominated by the time
it takes for many independent random walks
to meet each other.
Quite surprisingly,
we show that for a system below the percolation point
the broadcast time does not depend on the relation between
the mobility speed and the transmission radius.
In fact, we prove that
$\bt = \bigTht{n / \sqrt{k}}$
for any $0\leq r < r_c$,
even when the transmission range is significantly larger
than the mobility range in one step,
giving a tight characterization up to logarithmic factors.
Our result complements a recent result of
Peres et al.~(SODA 2011)
who showed that above the percolation point
the broadcast time is polylogarithmic in $k$.
\end{abstract}

\section{Introduction}

The emergence of mobile computing devices has added a new intriguing
component, \emph{mobility}, to the study of distributed systems.
In  fully mobile systems, such as wireless mobile ad-hoc networks (MANETs),
information is generated, transmitted and consumed within the mobile
nodes, and communication is carried out without the support of 
static structures such as cell towers.  These systems have been
implemented in vehicular networks and sensor networks attached to
soldiers on a battlefield or animals in a nature
reserve~\cite{OlariuW09, Gerla05, JuangOWMPR02, Stojmenovic02}.
Characterizing the power and limitations of mobile networks requires
new models and analytical tools that address the unique properties
of these systems~\cite{GrossglauserT02,ClementiPS09}, which include:
\begin{itemize}
\item
\emph{Small transmission radius}: the transmission range of individual agents
is restricted by limitations on energy consumption and interference
from other agents;
\item
\emph{Planarity}: agents reside, move and transmit on (subsets of) a
plane.  Low diameter graphs that are often used to model static
communication networks are not useful here;
\item
\emph{Dynamic communication graphs}: communication channels between
agents are changing dynamically as mobile agents move in and out of
the transmission radius of other agents;
\item
\emph{Relative speeds}: transmission speed is significantly faster
than the physical movement of the agents.  A message can execute several hops
before the network is altered by motion.
\end{itemize}

In this work we study the dynamics of information dissemination
between agents moving independently on a plane.  We consider a system
of $k$ mobile agents performing independent random walks on an
$n$-node grid, starting at time $0$ in a uniform distribution over the
grid nodes.  We focus on the fundamental communication primitive of
broadcasting a rumor originating at one arbitrary agent to all other 
agents in the system.  We characterize the \emph{broadcast time} $\bt$
of the system, which is the time it takes for all agents to receive
the rumor.

We model the spreading of information in a mobile system by a dynamic
communication graph process $\{G_t(r)~|~t\geq 0\}$, where the nodes of
$G_t(r)$ are the mobile agents, and two agents are connected by an
edge iff their distance at time $t$ is within their transmission
radius $r$.  We are interested in \newterm{sparse systems} in which
the transmission radius is below the percolation point $r_c \approx
\sqrt{n/k}$ \cite{Penrose03,PeresSSS11} (i.e., the minimum radius
which guarantees that $G_t(r_c)$ features a giant connected
component), and where, with high probability, no connected component
of $G_t(r_c)$ has more than a logarithmic number of agents. The
broadcast time in sparse systems is dominated by the time it takes for
many independent random walks to meet one another.  Incorporating the
fact that radio transmission is much faster than the motion of the
agents, we assume that the rumor can travel throughout a connected
component of $G_t$ within one step, before the graph is altered by the
motion.

Our main result is quite surprising: we show that below the
percolation point the broadcast time does not depend on the relation
between the mobility speed and the transmission radius. We prove that
$\bt = \bigTht{n / \sqrt{k}}$ for any $r$ below $r_c$, giving a tight
characterization up to logarithmic factors%
\footnote{The tilde notation hides polylogarithmic factors,
  e.g. $\bigOt{f(n)} = \bigO{f(n) \log^c n}$ for some constant $c$.}.
Our bound holds both when the transmission radius is significantly
larger than the mobility range (i.e., the distance an agent can travel
in one step), and when, in contrast to previous
work~\cite{ClementiMPS09, ClementiPS09}, the transmission radius as
well as the the mobility range are very small.  Our work complements a
recent result by Peres et al.~\cite{PeresSSS11} who proved an upper
bound polylogarithmic in $k$ for the broadcast time in a system of $k$
mobile agents which follow independent Brownian motions in $\R^d$, with
transmission radius above the percolation point.

Our analysis techniques are applicable to a number of interesting
related problems such as covering the grid with many random walks
and bounding the extinction time in random predator-prey systems.

\subsection{Related Work}

Information dissemination has been extensively studied in the
literature under a variety of scenarios and 
objectives. Due to space limitations, we restrict our attention to
the results more directly related to our work.

A prolific line of research has addressed broadcasting and gossiping
in static graphs, where the nodes of the graph represent active
entities which exchange messages along incident edges according to
specific protocols (e.g., \emph{push}, \emph{pull}, \emph{push-pull}).
The most recent results in this area relate the performance of the
protocols to expansion properties of the underlying topology, with
particular attention to the case of social networks, where
broadcasting is often referred to as \emph{rumor spreading}
\cite{ChierichettiLP10}.  (For a relatively recent, comprehensive
survey on this subject, see~\cite{HromkovicKPRU05}.)

Unfortunately, mobile networks do not feature
properties similar to those of social networks,
mostly because of the physical limitations
of both the movement and the radio transmission processes.
Indeed, as noted in~\cite{Kleinberg07},
the short range of communication attainable
by low-power antennas enforces the same local dynamics
that are typical of disease epidemics~\cite{Durrett99}
which requires physical proximity to propagate.
Indeed, the analysis of opportunistic networks,
where nodes relay messages as they come
close one to another, apply models 
from the study of human mobility~\cite{ChaintreauHCDGS07, Chaintreau08}.
Similarly, in the theory community there has been growing
interest in modeling and analyzing information dissemination in dynamic scenarios,
where a number of agents move either in a continuous space or along
the nodes of some underlying graph and exchange information when their
positions satisfy a specified proximity constraint.

In~\cite{ClementiMPS09,ClementiPS09} the authors study the time it takes
to broadcast information from one of $k$ mobile agents to all others.
The agents move on a square grid of $n$ nodes and in each time step,
an agent can (a) exchange information with all agents at distance at
most $R$ from it, and (b) move to any random node at distance at most
$\rho$ from its current position. The results in these papers only
apply to a very dense scenario where the number of agents is linear in
the number of grid nodes (i.e., $k=\bigTh{n}$).  They show that the
broadcast time is $\bigTh{\sqrt{n}/R}$ w.h.p., when $\rho = \bigO{R}$
and $R = \bigOm{\sqrt{\log n}}$ \cite{ClementiMPS09}, and it is
$\bigO{(\sqrt{n}/\rho)+\log n}$ w.h.p., when $\rho =
\bigOm{\max\{R,\sqrt{\log n}\}}$ \cite{ClementiPS09}.  These results
crucially rely on $R+\rho = \bigOm{\sqrt{\log n}}$, which implies that
the range of agents' communications or movements at each step defines
a connected graph.

In more realistic scenarios, like the one adopted in this paper, the
number of agents is decoupled from the number of locations (i.e., the
graph nodes) and a smoother dynamics is enforced by limiting agents to
move only between neighboring nodes. A reasonable model consists of
a set of multiple, simple random walks on a graph, one for each agent,
with communication between two agents occurring when they meet at the
same node.  One variant of this setting is the so-called \emph{Frog
  Model},
where initially one of $k$ agents is active (i.e., is performing a
random walk), while the remaining agents do not move.  Whenever an
active agent hits an inactive one, the latter is activated and starts
its own random walk.  This model was mostly studied in the infinite
grid focusing on the asymptotic (in time) shape of the set of vertices
containing all active agents~\cite{AlvesMP02, KestenS03}.

A model similar to our scenario is often employed to represent the
spreading of computer viruses in networks and the spreading time is
also referred to as \emph{infection time}.  Kesten and
Sidoravicius~\cite{KestenS05} characterized the rate at which an
infection spreads among particles
performing continuous-time random walks with the same jump
rate.  In \cite{DimitriouNS06}, the authors provide a general bound on
the average infection time when $k$ agents (one of them initially
affected by the virus) move in an $n$-node graph.  For general graphs,
this bound is $\bigO{t^* \log k}$, where $t^*$ denotes the maximum
average meeting time of two random walks on the graph, and the maximum
is taken over all pairs of starting locations of the random walks.
Also, in the paper tighter bounds are provided for the complete graph
and for expanders. Observe that the $\bigO{t^* \log k}$ bound
specializes to $\bigO{n \log n \log k}$ for the $n$-node grid by
applying the known bound on $t^*$ of \cite{AldousF98}.  A tight bound
of $\bigTh{(n \log n \log k)/k}$ on the infection time on the grid is
claimed in \cite{WangKK08}, based on a rather informal argument where
some unwarranted independence assumptions are made.  Our results show
that this latter bound is incorrect.

Recent work by Peres et al.~\cite{PeresSSS11} studies a process in
which agents follow independent Brownian motions in $\R^d$.  They
investigate several properties of the system, such as detection,
coverage and percolation times, and characterize them as functions of
the spatial density of the agents, which is assumed to be greater than
the percolation point.  Leveraging on these results, they show that
the broadcast time of a message is polylogarithmic in the number of
agents, under the assumption that a message spreads within a connected
component of the communication graph instantaneously, before the graph
is altered by agents' motion.

\subsection{Organization of the Paper}
The rest of the paper is organized as follows.  In
Section~\ref{sec:prelim}, we define the quantities of interest and
establish some technical facts which are used in the analysis.
Section~\ref{sec:gs} contains our main results: first, we prove the
upper bound on the broadcast time in the most restricted case, that
is, when the information exchange occurs through physical contact of
the agents (i.e., $r=0$), and then we provide a matching lower bound,
which holds for every value of the transmission radius $r$ below the
percolation point.  Finally, in Section~\ref{sec:conclusions} we
briefly discuss the connection between our result and other
interesting related problems and devise some future research directions.

\section{Preliminaries}
\label{sec:prelim}
In this paper, we study the dynamics of information exchange among a
set $A$ of $k$ agents performing independent random walks on an
$n$-node 2-dimensional square grid $\grid$,
which is commonly adopted as a discrete model
for the domain where agents wander.
We assume that $n \geq 2k$,
since sparse scenarios are the most interesting
from the point of view of applications;
however, our analysis can be easily extended to denser scenarios.
We suppose that the agents
are initially placed uniformly and independently at random on the grid
nodes. Time is discrete and agent moves are synchronized.  At each
step an agent residing on a node $v$ with $n_v$ neighbors ($n_v =
2,3,4$), moves to any such neighbor with probability $1/5$ and
stays on $v$ with probability $1-n_v/5$. With these probabilities it
is easy to see that at any time step the agents are placed
uniformly and independently at random on the grid nodes.
The following two lemmas contain important properties
of random walks on $\grid$, which will be employed for deriving our 
results\footnote{Throughout the paper, the distance between two
grid nodes $u$ and $v$, denoted by $\dist{u-v}$, is defined to be the
Manhattan distance.}. 
\begin{lemm}
\label{lemm:SRW}
Consider a random walk on $\grid$, starting at time $t=0$ at node $v_0$.
There exists a positive constant $c_1$ such that
for any node $v \neq v_0$,
\[
\prob{v \mbox{ is visited within } (\dist{v-v_0})^2 \mbox{ steps}}
\geq \frac{c_1}{\max \{1,\log (\dist{v-v_0})\}}.
\]
\end{lemm}
\begin{proof}
The Lemma is proved in \cite[Theorem~2.2]{AlvesMP02} for the infinite
grid $\Z^2$.  By the ``Reflection Principle''~\cite[Page 72]{Feller68},
for each walk in $\mathbb{Z}^2$ that started in
$\grid$, crossed a boundary and then crossed the boundary back to $\grid$,
there is a walk with the same probability that does not cross the
boundary and visits all the nodes in $\grid$ that were visited by the
first walk.  Thus, restricting the walks to $\grid$ can only change the
bound by a constant factor.
\end{proof}

\begin{lemm}
\label{lemm:props}
Consider the first $\ell$ steps of a random walk in $\grid$ which was at
node $v_0$ at time $0$.
\begin{enumerate}
\item\label{poin:dev} 
The probability that at any given step $1\leq i\leq \ell$ the random
walk is at distance at least $\geq \lambda\sqrt{\ell}$ from $v_0$ is
at most $2 e^{-\lambda^2/2}$.
\item\label{poin:range} 
There is a constant $c_2$ such that, with probability greater than
$1/2$, by time $\ell$ the walk has visited at least $c_2\ell/\log
{\ell}$ distinct nodes in $\grid$.
\end{enumerate}
\end{lemm}
\begin{proof}
We observe that the distance from $v_0$ in each coordinate defines a
martingale with bounded difference $1$.  Then, the first property
follows from the Azuma-Hoeffding Inequality~\cite[Theorem 2.6]{MitzenmacherU05}.
As for the second property, let $R_\ell$ be
the set of nodes reached by the walk in $\ell$ steps.  By
Lemma~\ref{lemm:SRW}, $\expe{R_\ell}=\bigOm{{\ell}/{\log \ell}}$ (even
when $v_0$ is near a boundary), while
$\vari{R_\ell}=\bigTh{{\ell^2}/{\log^4 \ell}}$
(see~\cite{Torney86}). The result follows by applying Chebyshev's
inequality.
\end{proof}

Let $M$ be a set of messages, which will be referred to as
\newterm{rumors} henceforth, such that for each $m \in M$ there is (at
least) one agent \newterm{informed} of $m$ at time $t=0$.  W.l.o.g.,
we can assume that the number of distinct rumors is at most equal to
the number of agent.  We denote by $\mes{a}{t}$ the set of rumors that
agent $a \in A$ is informed of at time $t$, for any $t \geq 0$;
possibly, $\mes{a}{0} = \emptyset$.  We assume that each agent is
equipped with a \newterm{transmission radius} $r \in \N$, representing
the maximum distance at which the agent can send information in a
single time step.

The spread of rumors can be represented by a dynamic communication
graph process \sloppy $\{G_t(r)~|~t\geq 0\}$, where $G_t(r)$, the
\newterm{visibility graph at time $t$}, is a graph with vertex set $A$
and such that there is an edge between two vertices iff the
corresponding agents are within distance $r$ at time $t$.
Following a
common assumption justified by the physical reality that the speed of
radio transmission is much faster than the motion of the
agents~\cite{PeresSSS11}, we suppose that rumors can travel throughout
a connected component of $G_t(r)$ before the graph is altered by the
motion.  We assume that within the same connected component agents
exchange all rumors they are informed of.  Formally, let $C$ be a
connected component of $G_t(r)$: for all $a \in C$, $\mes{a}{t} =
\bigcup_{a' \in C} \mes{a'}{t-1}$.  Note that the sets $\mes{a}{t}$
can only grow over time, that is, agents do not ``forget'' rumors.
The following quantities will be studied in this paper.
\begin{defi}[Broadcast Time, Gossip Time]
The \newterm{broadcast time} $\bt^m$ of a rumor $m \in M$ is the
first time at which every agent is informed of $m$, that is, for all
$t \geq \bt^m$ and $a \in A$, $m \in \mes{a}{t}$.  The
\newterm{gossip time} $\gt$ of the system is the first time at
which every agent is informed of every rumor, that is, for any $t
\geq \gt$ and $a \in A$, $\mes{a}{t} = M$.
\end{defi}

Note that both $\bt^m$ and $\gt$ depend on the transmission radius
$r$, but we will omit this dependence to simplify the notation.
We will also write $\bt$ instead of $\bt^m$ when the message $m$
is clearly identified by the context.

\section{Broadcasting Below the Percolation Point}
\label{sec:gs}

In this section we give bounds to the broadcast time $\bt$ of a
rumor when the transmission radius is below the percolation point $r_c
\approx \sqrt{n/k}$, that is, when all the connected components of
$G_t(r)$ comprise at most a logarithmic number of agents.  In this
regime, we show that quite surprisingly $\bt$ does not depend on the
relation between the mobility speed and the transmission radius, the
reason being that the broadcast time is dominated by the time it takes
for many independent random walks to intersect one another.  In
Subsection~\ref{sec:gsub} we prove an upper bound on the broadcast
time $\bt$ in the extreme case $r=0$, that is, when agents can
exchange information only when they meet on a grid node. The same
upper bound clearly holds for any other $r>0$. Then, in
Subsection~\ref{sec:gslb} we show that the upper bound is tight,
within logarithmic factors, for all values of the transmission radius
below the percolation point.  We also argue that the bounds on $\bt$
easily extend to gossip time $\gt$.

\subsection{Upper Bound on $\bt$}
\label{sec:gsub}

The main technical ingredient of the analysis carried out in this
subsection is the following lower bound on the probability that two
random walks $\bar{a}, \bar{b}$ on the grid meet within a given time
interval and not too far from their starting positions, which is a
result of independent interest.  Observe that considering the
difference random walk $\bar{a} - \bar{b}$ and computing the
probability that it hits the origin in the prescribed number of steps
does not provide any information about the place where the meeting
occurs, hence it is not immediate to derive our result through that
approach.

\begin{lemm}
\label{lemm:MeetingProbability}
Consider two independent simple random walks on the grid
$\bar{a} = \lis{a_0, a_1, \ldots}$, and $\bar{b} = \lis{b_0, b_1, \ldots}$,
where $a_t$ and $b_t$ denote the locations of the walks at time $t \geq 0$.
Let $d=\dist{a_0 -b_0}\geq 1$ and define $D$ to be the set of nodes at
distance at most $d$ from both $a_0$ and $b_0$.
For $T=d^2$, there exists a constant $c_3 > 0$ such that
\[
	P_{\bar{a},\bar{b}}(T) \triangleq \prob{\exists t\leq T \text{ such that } a_t=b_t \in D} \geq \frac{c_3}{\max\{1, \log d\}}.
\]
\end{lemm}
\begin{proof}
The case $d=1$ is immediate. Consider now the case $d>1$.  Let $P_t
(w, x)$ denote the probability that a walk that started at node $w$ at
time 0 is at node $x$ at time $t$, and let $R(w,u,D,s)$ be
the expected number of times that two walks which started at nodes $w$
and $u$ at time 0 meet at nodes of $D$ during the time
interval $[0,s]$, then
\[
R(w,u,D,s)=\sum_{t=0}^s \sum_{x \in D} P_t (w, x)P_t (u,x).
\]
Let $\tau (a,b)$ be the first meeting time of the walks $\bar{a}$ and
$\bar{b}$ at a node of $D$. Then
\[
R(a_0,b_0,D,T) = 
\sum_{t=0}^T 
\prob{\tau (a,b)=t}R(a_t,a_t,D,T-t) \leq 
P_{\bar{a},\bar{b}}(T) \max_x R(x,x,D,T).
\]
Thus, 
\[
P_{\bar{a},\bar{b}}(T) \geq 
\frac{R(a_0,b_0,D,T)}{\max_x R(x,x,D,T)}.
\]
It is easy to verify that $\card{D} \geq d^2 / 4$.
Applying Theorem 1.2.1 in \cite{Lawler91} we have:
\begin{eqnarray*}
R(a_0,b_0,D,T)
	&\geq &\sum_{t=0}^{T} \sum_{x\in D} P_t (a_0, x)P_t (b_0,x)\\
	&\geq &\sum_{t=\frac{T}{2}+1}^{T} \sum_{x\in D} 4\left(\frac{1}{\pi t}\right)^2e^{-\frac{\dist{x-a_0}^2 + \dist{x-b_0}^2}{t}}.
\end{eqnarray*}
By bounding $\dist{x-a_0}^2$ and $\dist{x-b_0}^2$ from above with $T$
in the formula, easy calculations show that $R(a_0,b_0,D,T) =
\bigOm{1}$.  Similarly, using the fact that there are no more than
$4i$ nodes at distance exactly $i$ from $x$, we have:
\begin{eqnarray*}
\max_x R(x,x,D,T) &\leq& 1 + \sum_{t=1}^T \sum_{i = 1}^{t} 4i\, 4\left(\frac{1}{\pi t}\right)^22 e^{-\frac{i^2}{t}} \\
&\leq & 1 + \left(\frac{4}{\pi}\right)^2 \sum_{t=1}^T \frac{1}{t^2} \left(\left(\sum_{i=1}^{\sqrt{t}} i\right) + \left(\sum_{i=1+\sqrt{t}}^{t} i e^{-i^2/t}\right)\right)\\
&\leq & 1 + \left(\frac{4}{\pi}\right)^2 \sum_{t=1}^T \frac{1}{t^2} \left(\frac{t}{2} + \left(\sum_{i=1+\sqrt{t}}^{t} i^2 e^{-i^2/t}\right)\right)\\
&\leq & 1 + \left(\frac{4}{\pi}\right)^2 \sum_{t=1}^T \frac{1}{t^2} \left(\frac{t}{2} + \frac{e}{(e-1)^2} t \right) = \bigO{\log T}.
\end{eqnarray*}

We conclude that there is a constant $c_3>0$ such that
$P_{\bar{a},\bar{b}}(T) \geq c_3/\log d$.\qedhere
\end{proof}

The reminder of this section is devoted to proving the following upper
bound on the broadcast time of a single rumor $m$ in the case $r=0$.
We assume that  $\mes{a}{0} = \{m\}$ for some $a \in A$,
and $\mes{a'}{0} = \emptyset$ for any other $a' \neq a$.
\begin{theo}
\label{theo:UBSpreadingTime2}
Let $r = 0$. For any $k \geq 2$, with probability at least $1-1/n^2$,
\[
	\bt = \bigOt{\frac{n}{\sqrt{k}}}.
\]
\end{theo}

We observe that since the diameter of $\grid$ is $2\sqrt{n} - 2$,
we can use Lemma~\ref{lemm:MeetingProbability} to show that
with probability at least $1-1/n^2$, at time $8 n \log^2 n$
an agent  has met all other agents walking   in $\grid$.
Thus, the theorem trivially holds for $k = \bigO{\mbox{poly}\log(n)}$.

From now on we concentrate on the case $k=\bigOm{\log^3 n}$.
	\iffalse
	Albeit our reference scenario is defined with respect to a fixed number $k$
	of agents (the \newterm{exact model}), technically, it is easier to
	derive an upper bound on $\gt$ using a slightly modified model in
	which, initially, each grid node $v$ holds $K_v$ agents, where $K_v$
	is binomial random variable with distribution $B(k,1/n)$, and the
	$K_v$'s are mutually independent (the \newterm{binomial model}). We
	refer to $k/n = E[K_v]$ as the \newterm{density} of the binomial
	model, and let $\tilde{K} = \sum_v K_v$ be the random variable
	denoting the number of agents in a given instance of the model.  For
	sufficiently large density, namely $k = \bigOm{\log n}$, a standard
	argument based on Chernoff bound, shows that $\tilde{K}=\bigTh{k}$,
	with high probability.  Then, by
	\cite[Corollary~5.9]{MitzenmacherU05}, a high-probability result for
	the binomial model with $k = \bigOm{\log n}$ implies a similar
	high-probability result in the exact model with $k$ agents.
	
	The argument proceeds as follows.
	\fi
We tessellate $\grid$ into
\newterm{cells} of side $\ell \triangleq \sqrt{{14 n \log^3 n}/{(c_3 k){}}}$,
where $c_3$ is defined in Lemma~\ref{lemm:MeetingProbability}.
We say that a cell $Q$ is
\newterm{reached} at time $t_Q$ if $t_Q$ is the first time when a node
of the cell hosts an agent informed of the rumor and we call this first
visitor the \newterm{explorer} of $Q$.
We first show that, after a suitably
chosen number $T_1 = \bigO{\ell^2 \log^4 n}$ of steps past $t_Q$,
there is a large number of informed agents 
within distance $\bigO{\ell \log^{5/2} n}$ from $Q$.
Furthermore, we show that while the rumor spreads to cells
adjacent to $Q$, at any time $t \geq t_Q + T_1$ a large number of
informed agents are at locations close to $Q$.
These facts will imply that
the exploration process proceeds smoothly and that all agents are
informed of the rumor shortly after all cells are reached.

The above argument is made rigorous in the following sequence of
lemmas.
\begin{lemm}
\label{lemm:FirstPhase}
Consider an arbitrary $\ell \times \ell$ cell $Q$ of the tessellation.
Let $T_1 = 16 \beta \gamma \ell^2 \log^4 n$ and
$c_4 = 8\sqrt{5\beta\gamma}$,
where $\beta = 7/(2c_1)$ and $\gamma = 18/c_3$.
By time $\tau_1 = t_Q + T_1$, at least $4\beta \log^2 n$ agents are informed and are
at distance at most $2 (1 + c_4 \log^{5/2} n) \ell$ from $Q$,
with probability $1-{1/n^8}$, for sufficiently large $n$.
\end{lemm}
\begin{proof}
Since at any given time the agents are at random and independent
locations, by the Chernoff bound we have that the following
\emph{density condition} holds with probability at least $1-1/n^9$,
for sufficiently large $n$:
for any cell $Q'$ and any time instant $t \in [0, n\log^4 n]$,
the number of agents residing in cell $Q'$ at time $t$
is at least $(7 \log^3 n)/{c_3}$.
In the rest of the proof, we assume that the density condition holds.

First, we prove that, by time $\tau_1$, there are at least $4\beta
\log^2 n$ informed agents in the system. We assume that at every time
step $t \in [t_Q,\tau_1]$ there is always an uninformed agent in the
same cell where the explorer resides (otherwise the sought property
follows immediately by the density condition). For $1 \leq i \leq
4\beta \log^2 n$, let $t_i \geq t_Q$ be the time at which the explorer
of $Q$ informs the $i$-th agent.  For notational convenience, we let
$t_0 = t_Q$.  To upper bound $t_i$, for $i > 0$, we consider a
sequence of $\gamma \log^2 n$ consecutive, non-overlapping time
intervals of length $4 \ell^2$ beginning from time $t_{i-1}$.  By the
previous assumption, at the beginning of each interval the cell where
the explorer resides contains an uninformed agent $a$.  Hence, by
Lemma~\ref{lemm:MeetingProbability}, the probability that the explorer
fails to meet an uninformed agent during all of these intervals is
\[
	\prob{t_i >  t_{i-1} + 4\gamma \ell^2 \log^2 n}
		\leq \left(1-c_3/\log(2\ell)\right)^{\gamma \log^2 n}
		\leq 1/n^9,
\]
where the last inequality holds for sufficiently large $n$
by our choice of $\gamma$.
By iterating the argument
for every $i$, we conclude that with probability at least $1 - 4\beta \log^2 n/n^9$,
there are at least $4\beta \log^2 n$ informed agents at time
$\tau_1$.  Let $I$ denote the set of informed agents identified
through the above argument, and observe that each agent of $I$ was in
the cell containing the explorer at some time step $t \in [t_Q, \tau_1]$.

To conclude the proof of the lemma, we note that, by
Lemma~\ref{lemm:props}, the probability that
the explorer, during the interval $[t_Q, \tau_1]$,
reaches a grid node at distance greater than $(c_4 \log^{5/2} n) \ell$
from its position at time $t_Q$ is bounded by $2 T_1 / n^{10}$.
Consider an arbitrary agent $a \in I$. As observed above,
there must have been a time instant $\bar{t} \in [t_Q, \tau_1]$ when
$a$ and the explorer were in the same cell, hence at distance at most
$(2 + c_4 \log^{5/2} n) \ell$ from $Q$.  From time $\bar{t}$ until
time $\tau_1$ the random walk of agent $a$ proceeds independently for
the random walk of the explorer.  By applying again
Lemma~\ref{lemm:props}, we can conclude that the probability that one
of the agents of $I$ is at distance greater than $2 (1 + c_4 \log^{5/2} n) \ell$
from $Q$ at time $\tau_1$ is at most $8 \beta \log^2 n / n^9$.
By adding up the upper bounds to the probabilities that the event stated
in the lemma does not hold, we get
$1/n^9 + 4\beta \log^2 n/n^9 + 2 T_1 / n^{10} + 8 \beta \log^2 n / n^9$,
which is less than $1/n^8$ for sufficiently large $n$.
\end{proof}

\begin{lemm}
\label{lemm:SecondPhase}
Consider an arbitrary $\ell \times \ell$ cell $Q$ of the tessellation.
Let $T_1, \tau_1, c_4$ and $\beta$ be defined as in
Lemma~\ref{lemm:FirstPhase}, and let
$T_2 = (2(2 + c_4 \log^{5/2} n)\ell)^2 $,
$\tau_2 = \tau_1 + T_2$,
and $c_5 = (4 \sqrt{\log 16}) c_4$.
Then, the following two properties hold with probability
at least $1-1/n^6$ for $n$ sufficiently large:
\begin{enumerate}
\item\label{point:re-exploration}
For $Q$ and for each of its adjacent cells,
there exists a time $t$, with $\tau_1 \leq t \leq \tau_2$,
at which there is an informed agent in the cell;
\item\label{point:re-population}
At any time $t$, with $\tau_1 \leq t \leq \tau_2 + T_1$, there are
at least $\beta \log^2 n$ informed agents at distance
at most $(2 + (2c_4 + c_5) \log^{5/2} n) \ell$ from $Q$.
\end{enumerate}
\end{lemm}
\begin{proof}
We condition on the event stated in Lemma~\ref{lemm:FirstPhase},
which occurs with probability $1-1/n^8$.
Hence, assume that by time $\tau_1$ there
are at least $4\beta \log^2 n$ informed agents at distance
at most $d_4 \triangleq 2 (1 + c_4 \log^{5/2} n) \ell$ from $Q$.
Consider the center node $v$ of $Q$ (resp., $Q'$ adjacent to $Q$),
so that at $\tau_1$ there are at least $4\beta \log^2 n$
informed agents at distance at most $d_4 + 2\ell$ from $v$.
By Lemma~\ref{lemm:SRW} the probability that $v$ is not touched by an
informed agent between $\tau_1$ and $\tau_2$ is at most
$\left(1-(c_1/\log (d_4 + 2\ell))\right)^{4\beta \log^2 n}$,
which is less than $1/n^7$, for sufficiently large $n$,
by our choice of $\beta$.
Thus, Point~\ref{point:re-exploration} follows.

As for Point~\ref{point:re-population}, consider an informed agent $a$ which,
at time $\tau_1$, is at a node $x$ at distance
at most $d_4$ from $Q$.
Fix a time $t \in [\tau_1, \tau_2 + T_1]$.
By Lemma~\ref{lemm:props} the probability that at time $t$ agent $a$ is
at distance greater than $(c_5 \log^{5/2} n) \ell$ from $x$ is at most $1/2$.
Hence, at time $t$ the average number of informed agents at distance
at most $d_4 + (c_5 \log^{5/2} n) \ell$ from $Q$ is at least $2\beta \log^2 n$.
Since agents move independently,
Point~\ref{point:re-population} follows by applying the
Chernoff bound to bound the probability that at time $t$ there are
less than $\beta \log^2 n$ informed agents at distance
at most $d_4 + (c_5 \log^{5/2} n) \ell$ from $Q$, and by applying the union bound
over all time steps of the interval $[\tau_1, \tau_2 + T_1]$.
\end{proof}

We are now ready to prove the main theorem of this subsection:
\begin{proofof}{Theorem~\ref{theo:UBSpreadingTime2}}
As observed at the beginning of the subsection, we can limit ourselves
to the case $k=\bigOm{\log^3 n}$.  Consider the tessellation of
$\grid$ into $\ell \times \ell$ cells defined before, and
focus on a cell $Q$ reached for the first time at $t_Q$.
By Lemma~\ref{lemm:SecondPhase}, we know that with probability at
least $1-1/n^6$, in each time step $t \in [\tau_1, \tau_2 + T_1]$
there are at least $\beta \log^2 n$ informed agents at distance at
most $d_5 \triangleq (2 + (2c_4 + c_5) \log^{5/2} n) \ell$ from $Q$
and there exists a time $t' \in [\tau_1, \tau_2]$ such that an informed agent
is again inside $Q$.
By applying again the lemma, we can conclude that, with probability at
least $(1-1/n^6)^2$, at any time step $t'' \in [t'+T_1, t'+ 2 T_1 +
  T_2]$ there are at least $\beta \log^2 n$ informed agents at
distance at most $d_5$ from $Q$.  Note that the
two time intervals $[\tau_1, \tau_2 + T_1]$ and $[t'+T_1, t'+ 2 T_1 +
  T_2]$ overlap and the latter one ends at least
$T_1$ time steps later.  Thus, by applying the lemma $n \log^4 n$
times, we ensure that, with probability at least 
$(1-1/n^6)^{n \log^4 n} \geq 1-\log^4 n/n^5$, 
from time $\tau_1$ until the end of the broadcast, there
are always at least $\beta \log^2 n$ informed agents at distance at
most $d_5$ from $Q$.

Lemma~\ref{lemm:SecondPhase} shows that each of the neighboring cells
of $Q$ is reached within time $\tau_2 = t_Q + T_1 + T_2$ with
probability $1 - 1/n^6$. Therefore, all cells are reached within time
$T^* = ({2\sqrt{n}}/{\ell})(T_1 + T_2)$ with probability at least $1 - 1/n^5$.
Hence, by applying a union bound over all cells,
we can conclude that with probability at least 
$(1 - 1/n^5)(1-\log^4 n/n^4) \geq 1-1/n^3$
there are at least $\beta \log^2 n$ informed agents at
distance at most $d_5$ from each cell of the tessellation,
from time $T^*+T_1$ until the end of the broadcast.

Consider now an agent $a$ which, at time $T^*+T_1$, is uninformed and resides in
a certain cell $Q$.  By an argument similar to the one used to prove
Lemma~\ref{lemm:FirstPhase}, we can prove that $a$ meets at least one
of the informed agents around $Q$ within $\bigO{\ell^2 \log^5 n}$ time
steps with probability at least $1-1/n^6$. A union bound over all
uninformed agents completes the proof.
\end{proofof}

Observe that the broadcast time is a non-increasing function of the
transmission radius.  Therefore, the upper bound developed for the
case $r = 0$ holds for any $r > 0$, as stated in the following corollary.
\begin{coro}
\label{coro:monotone}
For any $k \geq 2$ and $r > 0$, $\bt = \bigOt{n/\sqrt{k}}$
with probability at least $1-1/n^2$.
\end{coro}

As another immediate corollary of the above theorem, we can prove that
the gossiping of multiple distinct rumors completes within the same
time bound, with high probability.
\begin{coro}
For any $k \geq 2$ and $r > 0$, $\gt = \bigOt{n/\sqrt{k}}$
with probability at least $1-1/n$.
\end{coro}

\subsection{Lower Bound on $\bt$}
\label{sec:gslb}

In this subsection we prove that the result of
Corollary~\ref{coro:monotone} is indeed tight, up to logarithmic
factors, for any value $r$ of the transmission radius below the
percolation point.  Note that this result is also a lower bound on
$\gt$ if there are multiple rumors in the system.  First observe that
with probability $1 - 2^{-(k-1)}$, there exists an agent placed at
distance at least $\sqrt{n}/2$ from the source of $m$.
W.l.o.g., we assume that the $x$-coordinates of the positions occupied
by such an agent and the source agent differ by at least $\sqrt{n}/4$
and that the latter is at the left of the former.  (The other cases
can be dealt with through an identical argument.)  In the proof, we
cannot solely rely on a distance-based argument since we need to take
into account the presence of ``many'' agents which may act as relay to
deliver the rumor.

We define the \emph{informed area} $\I(t)$ at time $t$ as the set of
grid nodes visited by any informed agent up to time $t$, and let
$\point{x}(t)$ to be the rightmost grid node in $\I(t)$.  The
\newterm{frontier} of $\I(t)$ is the border separating the informed
area from the remaining places of the grid.  We will show that there is
a sufficiently large value $T$ such that, at time $T$, there is at
least one uniformed agent right of $\point{x}(T)$.  We need the
following definition:
\begin{defi}[Island]
\label{defi:Island}
Let $A$ be the set of agents.  For any $\gamma > 0$, let $G_t(\gamma)$
be the graph with vertex set $A$ and such that there is an edge
between two vertices iff the corresponding agents are within distance
$\gamma$ at time $t$.
The \newterm{island} of parameter $\gamma$ of
an agent $a$ at time $t$ %
is the connected component of $G_t(\gamma)$ containing $a$.
\end{defi}
Next, we prove an upper bound on the size of the islands.
\begin{lemm}
\label{lemm:NoBigNeighborhood}
Let $\gamma = \sqrt{n / (4 e^6 k)}$.
Then, the probability that there exists
an island of parameter $\gamma$
in any time instant $0 \leq t \leq 8 n \log^2 n$
with more than $\log n$ agents is at most $1/n^2$.
\end{lemm}
\begin{proof}
Since at any given time the agents are uniformly distributed in $\grid$,
the probability that a given agent is within distance $\gamma$
of another given agent at time $t_0$ is bounded by $4 \gamma^2 / n$.
Fix a time $t_0$ and let $\B_w(t_0)$ denote the event that
there exists an island with at least $w>\log n$ elements at time $t_0$. 
Then, recalling that $w^{w-2}$ is the number of unrooted 
trees over $w$ labeled nodes, we have that 
\[
	\prob{\B_w(t_0)}
		\leq \binom{k}{w} w^{w-2} \left(\frac{4 \gamma^2}{n}\right)^{w-1}
		\leq \left(\frac{ek}{w}\right)^w w^{w-2} \left(\frac{4 \gamma^2}{n}\right)^{w-1}.
\]
Using definition of $\gamma$ and the bound $w \geq 1 + \log n$ and $k \leq n$, we have 
\[
	\prob{\B_w(t_0)} \leq \frac{ek}{w^2} e^{-5(w-1)}\leq \frac{e n}{w^2} \frac{1}{n^5}
		\leq \frac{1}{n^4},
\]
for a sufficiently large $n$.
Applying the union bound on $\bigO{n \log^2 n}$ time steps concludes the proof.
\end{proof}

Next we show that, with high probability,
the frontier of the informed area cannot advance too fast
if the transmission radius satisfies $r \leq \sqrt{n / (64 e^6 k)}$.
\begin{lemm}
\label{lemm:SlowFrontier}
Suppose $r \leq \sqrt{n / (64 e^6 k)}$.
Let $\gamma =  \sqrt{n / (4 e^6 k)}$ and let $t_0$
and $t_1 = t_0 + \gamma^2/(144 \log n)$ be two time steps.
Then, with probability $1 - 1/n^2$,
\[
	\dist{\point{x}(t_1) - \point{x}(t_0)} \leq (\gamma \log n) / 2.
\]
\end{lemm}
\begin{proof}
By Lemma~\ref{lemm:props}, with probability $1-2/n^3$ an agent cannot
cover a distance of more than $(\gamma - r) / 2$ in $\gamma^2/(144 \log n)$
time steps.  Thus, with probability $1-1/n^2$, up to time $t_1$ the
rumor cannot propagate (directly or through intermediate agents)
between islands. By applying Lemma~\ref{lemm:NoBigNeighborhood} we
conclude that at time $t_1$ the rightmost position touched by agents
of any island $I$ is at most $(\gamma \log n) / 2$ right of the
rightmost position occupied by agents of $I$ at $t_0$, which is not on
the right of $x(t_0)$. Thus, the lemma follows.
\end{proof}

Finally, we can prove the main theorem of the subsection:
\begin{theo}
\label{theo:LBSpreadingTime2}
For any $k \geq 2$, suppose $r \leq \sqrt{n / (64 e^6 k)}$.
Then, with probability \sloppy $1 - \max\{2^{-(k-1)}, 2/n^2\}$,
\[
	\bt = \bigOm{\frac{n}{\sqrt{k}\log^2 n}}.
\]
\end{theo}
\begin{proof}
As mentioned before, with probability $1 - 2^{-(k-1)}$
there exists an agent $a$ placed at
distance at least $\sqrt{n}/2$ from the source of the rumor and we may
assume that their $x$-coordinates differ by at least $\sqrt{n}/4$ and
that the uninformed agent is to the right of the source agent.
Let $T = n / (1152 e^3 \sqrt{k} \log^2 n)$ and
$\gamma = \sqrt{n / (4 e^6 k)}$.
By Lemma~\ref{lemm:SlowFrontier}, with probability $1 - 1/n$
the frontier cannot move right in $T$ steps more than
$(\gamma \log n / 2) T / (\gamma^2 / (144 \log n)) < \sqrt{n} / 8$.
By Lemma~\ref{lemm:props}, with probability $1 - 2/n^2$,
agent $a$ cannot move left more than $2 \sqrt{T\log n} < \sqrt{n}/8$,
so that agent $a$ cannot be informed by time $T$.
Hence, the broadcast time is at least
$\bt > T = \bigOm{n/(\sqrt{k}\log^2 n)}$
with probability $1 - \max\{2^{-(k-1)}, 2/n^2\}$.
\end{proof}

\section{Further Results and Future Research}
\label{sec:conclusions}

In this work we took a step toward a better understanding of the
dynamics of information spreading in mobile networks.  We proved a
tight bound (up to logarithmic factors) on the broadcast of a rumor in
a mobile network where agents perform independent random walks on a
grid and the transmission radius defines a system below the
percolation point.  Our result complements the work of Peres et
al.~\cite{PeresSSS11}, who studied the behavior of a similar system
above the percolation point.  A similar bound holds for the gossip
problem in this model, where at time $0$ each agent has a distinct
rumor and all agents need to receive all rumors.

Our analysis techniques are applicable to some interesting related
problems.  For example, similar bounds on the broadcast time $\bt$ and
the gossip time $\gt$ can be obtained for the Frog Model
\cite{AlvesMP02}, where only informed agents move and uninformed
agents remain at their initial positions.  In particular, we can show
that the broadcast time in the Frog Model is upper bounded by $\bt =
\bigOt{n/\sqrt{k}}$.  The argument is similar to the proof of
Theorem~\ref{theo:UBSpreadingTime2}, where
Lemma~\ref{lemm:MeetingProbability} is replaced with
Lemma~\ref{lemm:SRW} and the analysis of the initial phase of the
information dissemination process is carried out by using
Point~\ref{poin:range} of Lemma~\ref{lemm:props}. Also, a closer look to
Theorem~\ref{theo:LBSpreadingTime2} reveals that the same argument
employed in our dynamic model to bound $\bt$ (hence, $\gt$) from below 
applies to the Frog Model. Thus, we have tight bounds, up to
logarithmic factors, in this latter model as well.

Another measure of interest in systems of mobile agents is the
\newterm{coverage time} $\ct$, that is, the first time at which every
grid node has been visited at least once by an informed agent
\cite{PeresSSS11}. While in the Frog Model the broadcast time is
obviously upper bounded by the coverage time, this relation is not so
obvious in our dynamic model, since the coverage of the grid nodes
does not imply that all agents have been informed of the rumor.
Nevertheless, one can verify that, in our model, $\ct \approx \bt =
\bigOt{n/\sqrt{k}}$.  Indeed, by Point~\ref{point:re-population} of
Lemma~\ref{lemm:SecondPhase} and by Lemma~\ref{lemm:SRW}, after
$\bigO{\ell^2}$ steps from the first time at which an informed agent
reached a given cell, all the nodes of that cell have been visited by
some informed agent.  Hence, by the cell-by-cell spreading process
devised in the proof of Theorem~\ref{theo:UBSpreadingTime2}, we can
conclude that the coverage time is bounded by $\bigOt{n/\sqrt{k}}$.
(In fact, the same tight relation between $\ct$ and $\bt$ can be
proved in the Frog Model.)

\sloppy Another by-product of our techniques is a high-probability upper bound
$\bigO{(n \log^2 n)/k + n \log n}$ on the
\newterm{cover time} of $k$ independent random walks on the $n$-grid
(i.e., the time until each grid node has been touched by at least one
such walk), improving on previous results~\cite{AlonAKKLT08,
  ElsasserS09} providing the same bound only for the expected
value. Finally, in a closely related scenario, namely a random
\newterm{predator-prey system} where $k = \bigOm{\log n}$ predators
are to catch moving preys on an $n$-node grid by performing independent
random walks~\cite{CooperFR09}, we can prove a
high-probability upper bound $\bigO{(n \log^2 n)/k}$ 
on the extinction time of the preys.

We are working now on extending our modeling and analysis techniques
to handle more complex planar domains that include both communication
and mobility barriers.

\noindent {\bf Acknowledgments:}
Many thanks to Jeff Steif for referring us to some crucial
references and to Andrea Clementi and Riccardo Silvestri for
pointing out a claim not fully justified in the proof of Lemma 4,
which appeared in the first version of the draft.

\phantomsection
\addcontentsline{toc}{chapter}{References}
\bibliographystyle{acm}
\bibliography{references}
\end{document}